\documentclass[12pt,preprint]{aastex}

\shorttitle{The Arches Cluster Mass Function}
\shortauthors{Stolte et al.}

\begin{document}

\title{The Arches Cluster - Evidence for a Truncated Mass Function?\footnotemark[1]}

\author{Andrea Stolte\altaffilmark{1}\altaffilmark{2}, Wolfgang Brandner\altaffilmark{1}, Eva K. Grebel\altaffilmark{3}\altaffilmark{1}, Rainer Lenzen\altaffilmark{1}, Anne-Marie Lagrange\altaffilmark{4}}
\altaffiltext{1}{Max-Planck-Institute for Astronomy, K\"onigstuhl 17, D-69117 Heidelberg, Germany,brandner@mpia.de,lenzen@mpia.de}
\altaffiltext{2}{University of Florida, 211 Bryant Space Science Center, Gainesville FL-32111-2055, USA,stolte@astro.ufl.edu}
\altaffiltext{3}{Astronomical Institute of the University of Basel, Venusstrasse 7, CH-4102 Binningen, Switzerland,grebel@astro.unibas.ch}
\altaffiltext{4}{Laboratoire d'Astrophysique de Grenoble, Universit\'e Joseph Fourier, F-38041 Grenoble Cedex 9, France,anne-marie.lagrange@cnrs-dir.fr}

\shorttitle{The Arches Cluster Mass Function}

\begin{abstract}
We have analyzed high-resolution, adaptive optics (AO) $HK$
observations of the Arches cluster obtained with NAOS/CONICA.
With a spatial resolution of 84 mas, the cluster center is uniquely 
resolved. From these data the present-day mass function (MF) of Arches 
is derived down to $\sim 4\,M_\odot$. 
The integrated MF as well as the core and $2^{nd}$ annulus MFs are consistent 
with a turn-over at $6-7\,M_\odot$.
This turn-over indicates severe depletion of intermediate
and low-mass stars in the Arches cluster, possibly caused by its 
evolution in the Galactic Center environment.
The Arches MF represents the first resolved observation
of a starburst cluster exhibiting a low-mass truncated MF.
This finding has severe implications for stellar population synthesis
modelling of extragalactic starbursts, the 
derivation of integrated properties such as the total mass of star clusters
in dense environments, the survival of low-mass remnants from starburst 
populations, and chemical enrichment during starburst phases.
\end{abstract}


\section{Introduction}

\footnotetext[1]{Based on NAOS/CONICA commissioning observations obtained at the ESO VLT on Paranal, Chile. \\{\sl send offprint requests to:} stolte@astro.ufl.edu}
The Arches Cluster is one of the rare compact starburst clusters 
forming in the Milky Way today. At a projected distance of only
25 pc from the Galactic Center (GC)
%, and estimated absolute GC distance
%between 50 and 90 pc \citep{PZ2002},
it forms a unique template for star and cluster formation in 
the nuclei of external galaxies. Even at an age of only 2 Myr, 
its elongated shape evidences the strong tidal forces acting
to disrupt the compact starburst cluster despite its large central density
of $3\times 10^5\,M_\odot {\rm pc^{-3}}$ \citep{Figer1999}.
N-body simulations suggest that clusters 
%exposed to the strong tidal forces 
in the GC environment will dissolve on 
timescales as short as 10 Myr \citep{Kim2000,PZ2002}. 
%In addition to 
%the forces of the external tidal field, the high central density of 
%the cluster causes short internal 
%dynamical timescales with a relaxation time of a few Myr
%as derived from simulations \citep{Kim2000} and estimated from the 
%observed MF \citep{Stolte2002}. The combined disruptive effects 
%of these forces 
This leaves us with very few clusters available to 
study nearby, spatially resolved starburst clusters, especially
at the Milky Way's present low cluster formation rate. 

The Arches cluster is known to be heavily mass segregated, with 
a flat present-day MF ($\Gamma \sim 0$ where the Salpeter slope is 
$\Gamma=-1.35$) in the cluster core ($r_{core}=0.2$ pc at a distance 
of 8 kpc) \citep{Stolte2002,Figer1999}. 
The concentration of high-mass stars in the cluster core
causes the integrated MF to be flattened
with respect to a Salpeter MF with a slope of 
$\Gamma \sim -0.9\ {\rm to} -1.1$ 
depending on the radial selection of the cluster area (Stolte 2003).
In this letter, we derive the present-day MF of the Arches 
cluster from AO data with a spatial resolution of $0.\!\!^{''}084$.
This comprises the currently deepest photometry of the cluster
core with the highest spatial resolution, which allows us to derive 
the present-day MF down to $\sim 5\,M_\odot$ in the cluster core 
($r < 0.2$ pc) and $2\,M_\odot$ at larger radii. 

\section{Observations and data reduction}

NAOS-CONICA (NACO) is the ESO VLT facility AO
system with its infrared camera. NAOS is a 189 element Shack-Hartmann AO
system with two wavefront sensors operating in the visual and near-infrared 
\citep{Rousset2003}. CONICA is a 1 to 5 micron imaging camera and spectrograph
designed for diffraction limited observations with the VLT 8m mirror \citep{Lenzen2003}.

\begin{figure}
\plotone{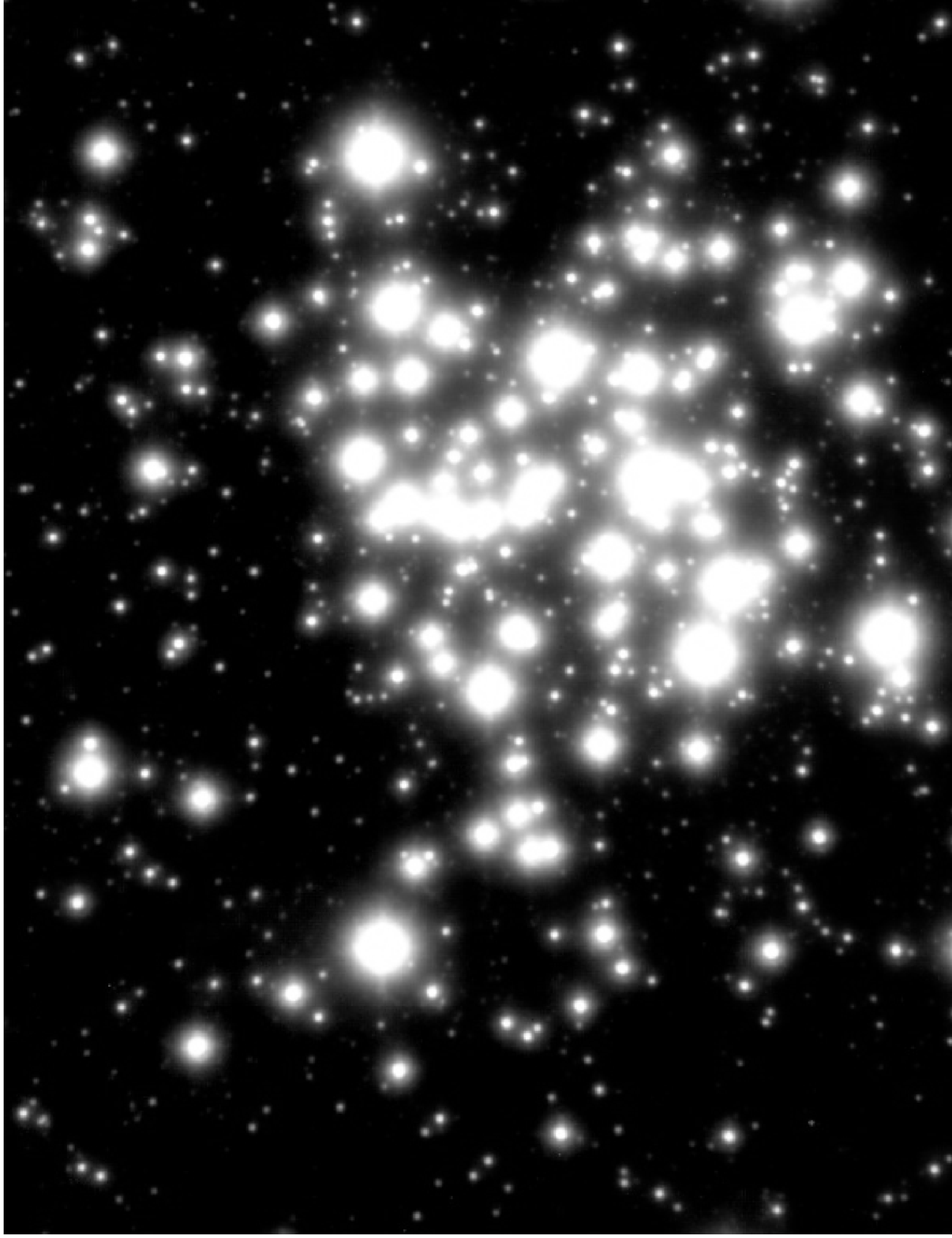}
\caption{\label{hkfaint} 
NACO combined $HK$ image of the Arches cluster. 
The field of view is $24^{''} \times 24^{''}$ with a 
final resolution of $0.\!\!^{''}084$. North is up, East left.}
\end{figure}

The Arches cluster ($\alpha =17h45m50s$, $\delta =-28^{o} 49^{'} 28^{''}$ (J2000)) 
was observed during the NACO commissioning phase in March 2002
to probe the AO crowded field performance. 
The medium resolution camera
($0.\!\!^{''}027$/pixel) optimised for $K$-band observations was used,
covering a field of view of 27$^{''}$. A foreground star with $V \sim 16$ mag
close to the cluster served as the guide source. Moderate Strehl ratios of
14\,\% in $H$-band and 20\,\% in $K$-band were achieved, resulting in a 
resolution of $0.\!\!^{''}084$ in both filters. 
%A combination of long
%and short exposures was used to increase the dynamic range. 
Integration
times were 14 min in $H$ and 7 min in $K_s$. The detection
limits on the combined images shown in Fig.~\ref{hkfaint}
were $K_s < 20.5$ and $H < 22$ mag.

Standard data reduction 
was performed under IRAF, images were combined using {\sl drizzle}
with $2\times 2$ oversampling to enhance source detection, and 
PSF fitting photometry was obtained with DAOPHOT.
The isoplanatic patch was large during the observations, such that no
elongation of the PSF at larger distances 
from the guide star due to anisoplanatism is observed. 
%The photometric performance on the cluster center is limited most critically
%by the high stellar density. 
%As isolated PSF stars are rare, 
%a constant PSF was created from a few isolated stars after iterative 
%neighbour subtraction.
HST/NICMOS observations of the Arches cluster \citep{Figer1999} were used 
to probe the stability of the photometric correction 
and derive a uniform photometric calibration over the NACO field. 
The calibration procedure is described in detail with respect to the 
Gemini/Hokupa'a data set in Stolte et al.~2002.
In the case of the NACO data, the zeropoint was constant and no color terms 
were found. In the following, all magnitudes 
are shown in the HST/NICMOS system, with $K$ corresponding to the m205 NICMOS
broadband filter, and $H$ to m160. All isochrones are transformed into 
the NICMOS system accordingly \citep{Brandner2001}.

\subsection{Artificial star simulations}

The completeness in the crowding limited cluster field was derived
from the recovery fraction of artificial stars. In order to avoid
changes in the local stellar density, only 200 stars were inserted
per frame on a total of 100 frames, or 20000
stars. The same stars randomly inserted into the $K$-band frame were
transformed to instrumental $H$-band magnitudes using the characteristic
Arches main sequence color of $H-K = 1.63$ mag. 
This is important because only stars detected in both $H$ and $K$
simultaneously enter the MF calculation. PSF photometry was derived
in the same way as the original photometry.
Recovered stars were selected by uncertainty and color in the same way as 
real stars entering the MF. The recovery fraction is determined for each 
annulus individually, such that increasing crowding effects at smaller radii
are taken into account. 
In the cluster core, the stellar density is significantly higher
than in the outer annuli. The simulations showed that inserted bright stars 
influence the recovery fraction of fainter stars severely. Therefore, 
the core field with the highest stellar density was excerpted from the NACO field.
On this excerpt, only 10 stars within bins of 1 mag were added on 100 artificial 
frames, leading to 1000 stars per magnitude bin. 
For the core number counts ($r < 0.2$ pc), only the recovery fraction from the 
core simulation was used. 

\section{The Arches present-day MF}
\label{nacomfsec}

\begin{figure}
\epsscale{0.75}
\plotone{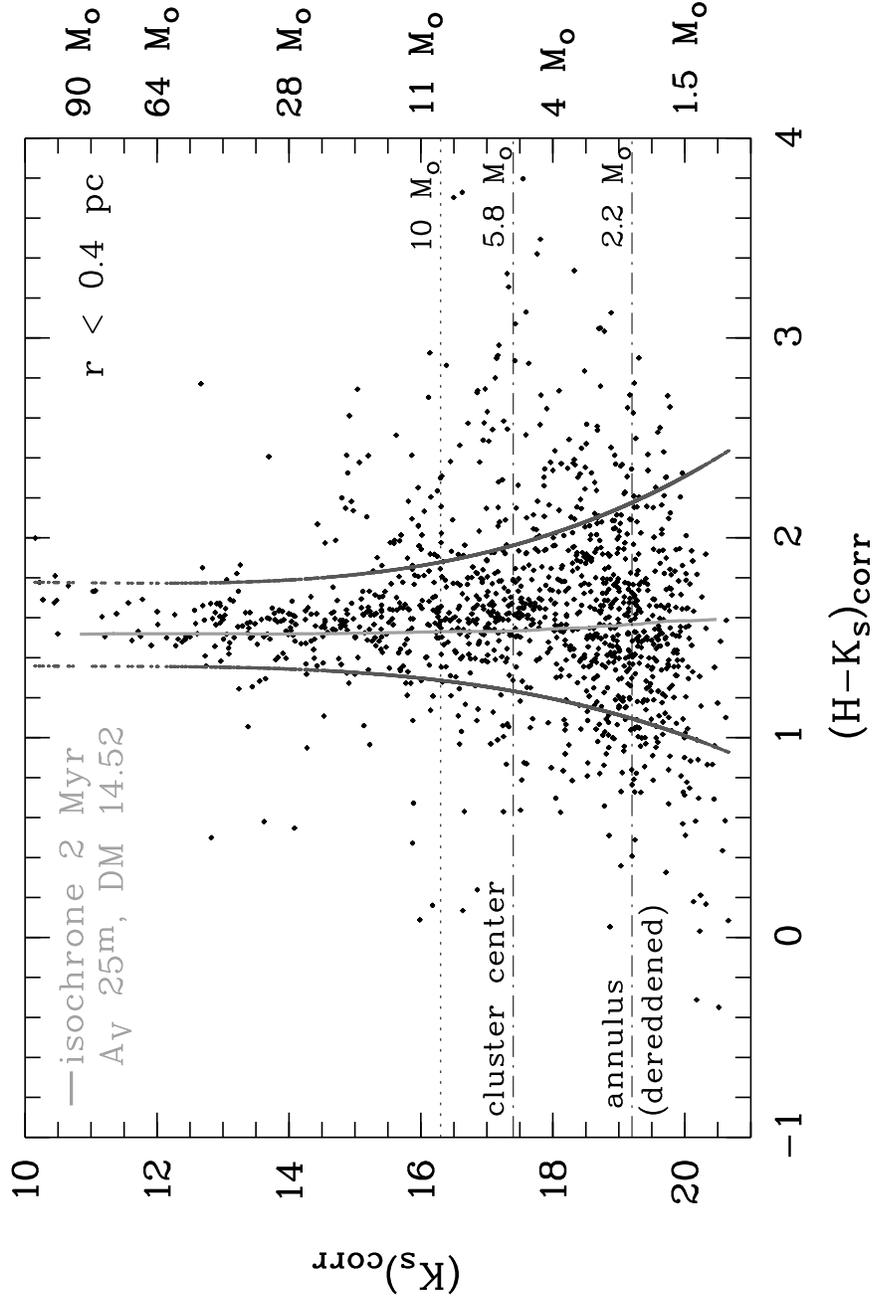}
\caption{\label{nacocmd}
Differential extinction corrected CMD of the Arches cluster. 
The 2 Myr Geneva isochrone used to derive the MF is shown as the solid
central line. The color cut used 
to reject foreground and background contamination (enveloping lines) widens
according to the increasing photometric uncertainty at fainter magnitudes.
The 50\% incompleteness limits in the cluster core and the $2^{nd}$ annulus
($0.2 < r < 0.4$ pc) are shown as 
dashed lines, and masses are labeled as derived from the isochrone displayed.
The radial selection to $r < 10^{''}$ or 0.4 pc minimizes field star contamination.}
\end{figure}

Prior to the mass determination, stellar magnitudes and colors were
corrected for the systematic variation in extinction observed over 
the Arches field \citep{Stolte2002}. 
Foreground and background sources were rejected by applying a color cut
as shown in Fig.~\ref{nacocmd}. The increasing photometric uncertainty
towards fainter stars as indicated by artificial star tests was taken 
into account by widening the color selection towards fainter magnitudes. 
Due to the location in the dense GC region, field contamination increases 
towards fainter magnitudes. This is apparent in the increase
of blue foreground and red background sources in the CMD below $K=15.5$ mag
($14\,M_\odot$),
%, where the density of blue foreground as well as red background sources 
%becomes noticeable, 
while contamination at higher masses is negligible. 
A dense concentration of bulge stars can be seen at the faint end,
($K>18$ mag), which is beyond the 50\% completeness limit in the MF and does 
not affect the MF fit. Field contamination becomes particularly important beyond
a radius of 10$^{''}$. As no background field is available, 
the MF is limited to radii $r < 10^{''}$ or 0.4 pc, twice the core radius of 
$0.2$ pc \citep{Stolte2002} to minimize field contamination.
As some field star component is still present in the radial and color selection, 
this results in a conservative upper limit to the flattened MF. 

The present-day MF was derived from the corrected CMD (Fig.~\ref{nacocmd})
converting $K$-band magnitudes into masses using a 2 Myr Geneva 
main sequence isochrone \citep{Lejeune2001} with solar metallicity
\footnote{The metallicity in the high-mass component of Arches was recently 
found to be solar\citep{Najarro2004}, while supersolar metallicity isochrones were
used in earlier studies, see discussion in Figer et al.~1999, Stolte et al.~2002
and Stolte 2003.}.
A distance modulus of 14.52 mag and foreground extinction
of $A_V=25.2$ mag were applied to the isochrone, and reddened luminosities were
transformed into the NICMOS system using color equations derived from 
comparison of five NICMOS standards and reddened stars in the Arches field 
with 2MASS. %($m205-K_s = (0.19\pm 0.04)\cdot (H-K_s) + (0.02 \pm 0.03)$, 
%$m160-m205 = (0.93\pm 0.05)\cdot (H-K_s) + (0.08 \pm 0.03)$).
Present-day masses were used, as present-day properties are calibrated 
against observations while the re-calculation to initial masses is heavily 
dependent on the mass-loss model. No correction was made for stellar 
evolution, which affects only stars above $50\,M_\odot$ ($\log M/M_\odot \geq 1.7$), 
to avoid uncertainties in stellar evolution modelling of very massive stars.

\begin{figure}
\plotone{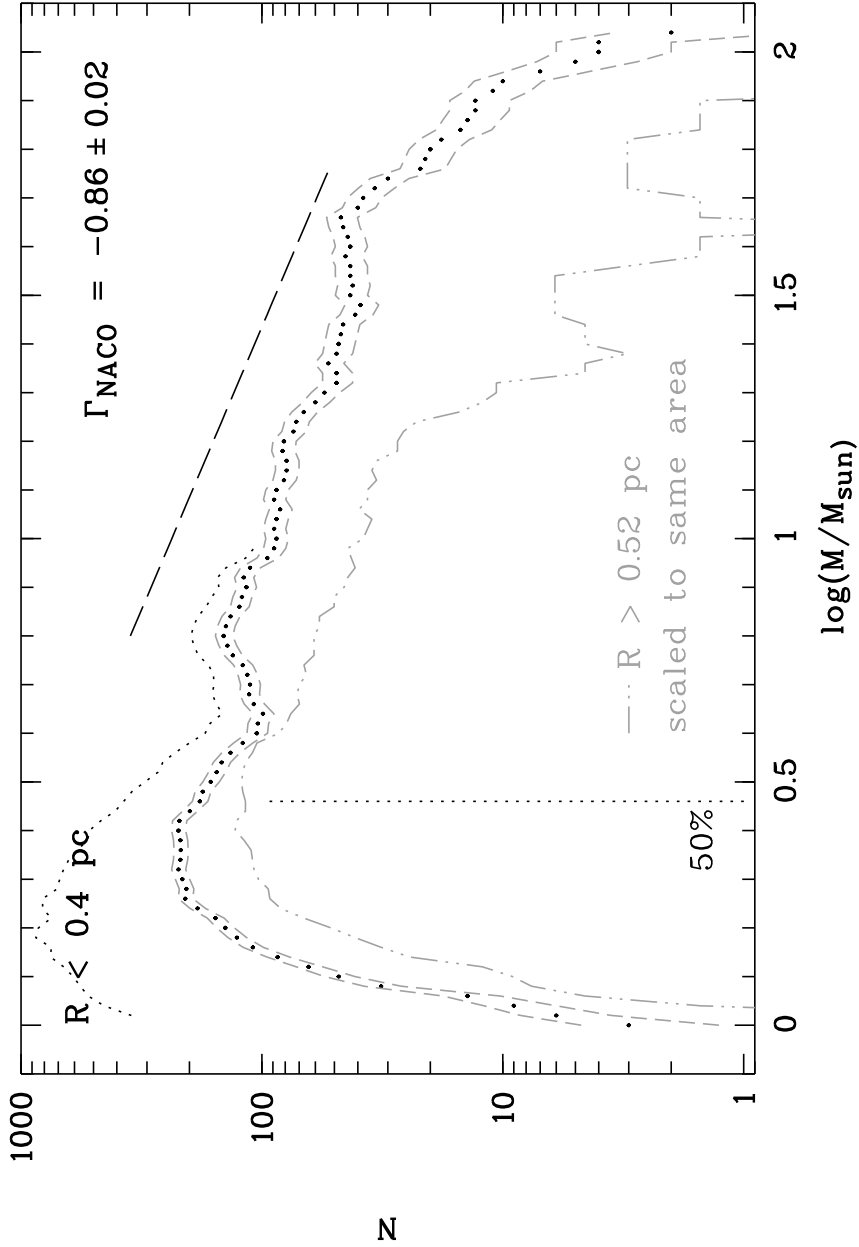}
\epsscale{1.0}
\caption{\label{nacomf}
Present-day MF as derived from a 2 Myr Geneva isochrone applied to 
the CMD shown in Fig.~\ref{nacocmd}. A bin width of $\Delta\log M/M_\odot = 0.2$ and 
bin shift of $\delta\log M/M_\odot = 0.02$ was used to obtain individual 
number counts. The gray enveloping lines denote the Poisson uncertainty for
each point. The slopes and fitting uncertainties are derived from a weighted, 
linear least-squares fit between $6 < M < 60\,M_\odot$ (dashed line).
The high-mass MF is with $\Gamma=-0.86$ moderately flattened with respect 
to a Salpeter IMF ($\Gamma=-1.35$), indicating a bias to high-mass objects,
and displays a TO at $\log M/M_\odot=0.8$, $M=6\,M_\odot$.  
An upper limit to the possible field star contamination (dash-dotted line) 
derived from the outer NACO field, indicates that field stars are dominant 
below $\log m =0.6$ or $4\,M_\odot$.}
\end{figure}

Stars were counted in magnitude intervalls of $\Delta\log M/M_\odot = 0.2$.
In order to minimize binning effects, ten MFs were calculated with bins shifted by 
$\delta\log m = 0.02$. 
The points from all ten calculations are included to yield the present-day
MF presented in Fig.~\ref{nacomf}. The advantage of this method lies in the fact that
features in the MF are not erased as a consequence of binning, and that the resultant
MF fit is statistically more significant. A linear least squares fit was performed
to yield the slopes. The true uncertainty in the derived MF slope is larger than 
the formal fit uncertainty, and 
is dominated by the choice of the isochrone, the photometric uncertainties of individual 
stars, and individual extinction. 
We estimate the true uncertainty to $\Gamma\ \pm 0.15$ from experimenting
with different fitting ranges and isochrone parameters (metallicity, age; see Stolte 2003 
for details). 

The most striking feature in the present-day MF of Arches (Fig.~\ref{nacomf}) is 
a change in the sign of the MF slope at $\sim 6\,M_\odot$ indicative of a turn-over (TO).
The MF has therefore been fitted in the mass range 
$6 < M < 65\,M_\odot$ ($\log M/M_\odot > 0.8$).
The MF is with a slope of $\Gamma = -0.86$ moderately flattened with respect
to a Salpeter value of $\Gamma=-1.35$. This slope is in very good agreement
with previous values of $\Gamma = -0.77$ derived from HST/NICMOS data 
\citep{Figer1999,Stolte2002}, and $\Gamma = -0.82$ 
derived from Gemini/Hokupa'a data (Stolte 2003). Below $M < 6\,M_\odot$ 
($\log M/M_\odot < 0.8$) the MF deviates from a power-law, but appears 
to {\sl decrease} towards lower masses between $0.6 < log M/M_\odot < 0.8$.
Around $4\,M_\odot$, the increase in the apparent MF is likely caused by the
increasing contribution of field stars. 

As field contamination increases with 
decreasing mass especially for masses below $\sim 10\,M_\odot$ ($K=16.3$ mag),
the observed TO will be enhanced when field stars are subtracted.
As an upper limit to the field star contribution we consider stars
close to the edge of the NACO field, at cluster radii $r > 13^{''}$ (0.5 pc). 
The corresponding apparent MF, scaled to the same area (Fig.~\ref{nacomf}, 
dash-dotted line), suggests that the population
is dominated by field stars or segregated low-mass cluster members below 
$3\,M_\odot$, while the high-mass end above $15\,M_\odot$ is entirely dominated 
by cluster stars. Between 3 and $15\,M_\odot$, the range where the TO 
occurs, field stars may contribute as much as 50\%. The discrepancy 
between the high- and the intermediate mass regime in the Arches MF will therefore
become more pronounced as soon as proper field star estimates are available.
The TO in the MF is supported by the fact that no evidence for such a 
feature is observed in the outer field population, as would be expected if 
field star contamination introduces this feature into the MF. 

Two possibilities can be envisioned to explain the flattening in the Arches MF.
The MF can either be truncated at the low-mass end, which implies that 
the dense GC environment has severe impacts on the local {\sl initial} MF (IMF),
or the tidal forces acting on the cluster caused rapid dynamical 
segregation on timescales as short as 2 Myr. Such dynamical evolution would
transport massive stars into the core and low-mass stars to larger radii.
The dynamical timescales for an Arches-like cluster are suggested to be 
as short as a few Myr from N-body simulations \citep{Kim2000,PZ2002}, 
such that dynamical disruption has likely
contributed to the observed spatial distribution of cluster 
stars. 
%Evidence for fast dynamical evaporation is observed in the 
%Quintuplet cluster, considered a twin of Arches, which appears less
%confined already at an age of 4 Myr \citep{Figer1999}.

\section{The MF turn-over in two annuli}
\label{nacocoresec}

\begin{figure*}
\plottwo{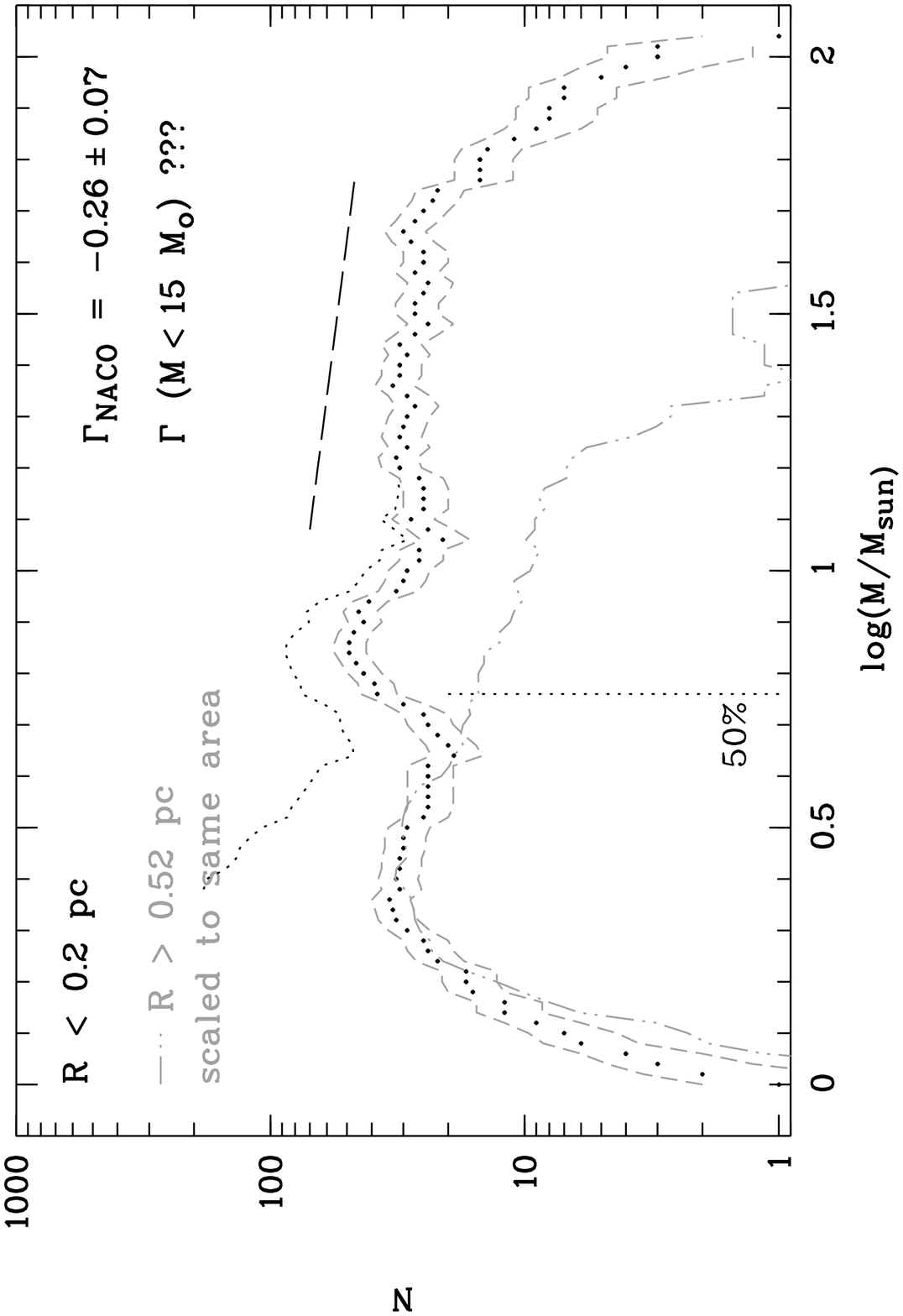}{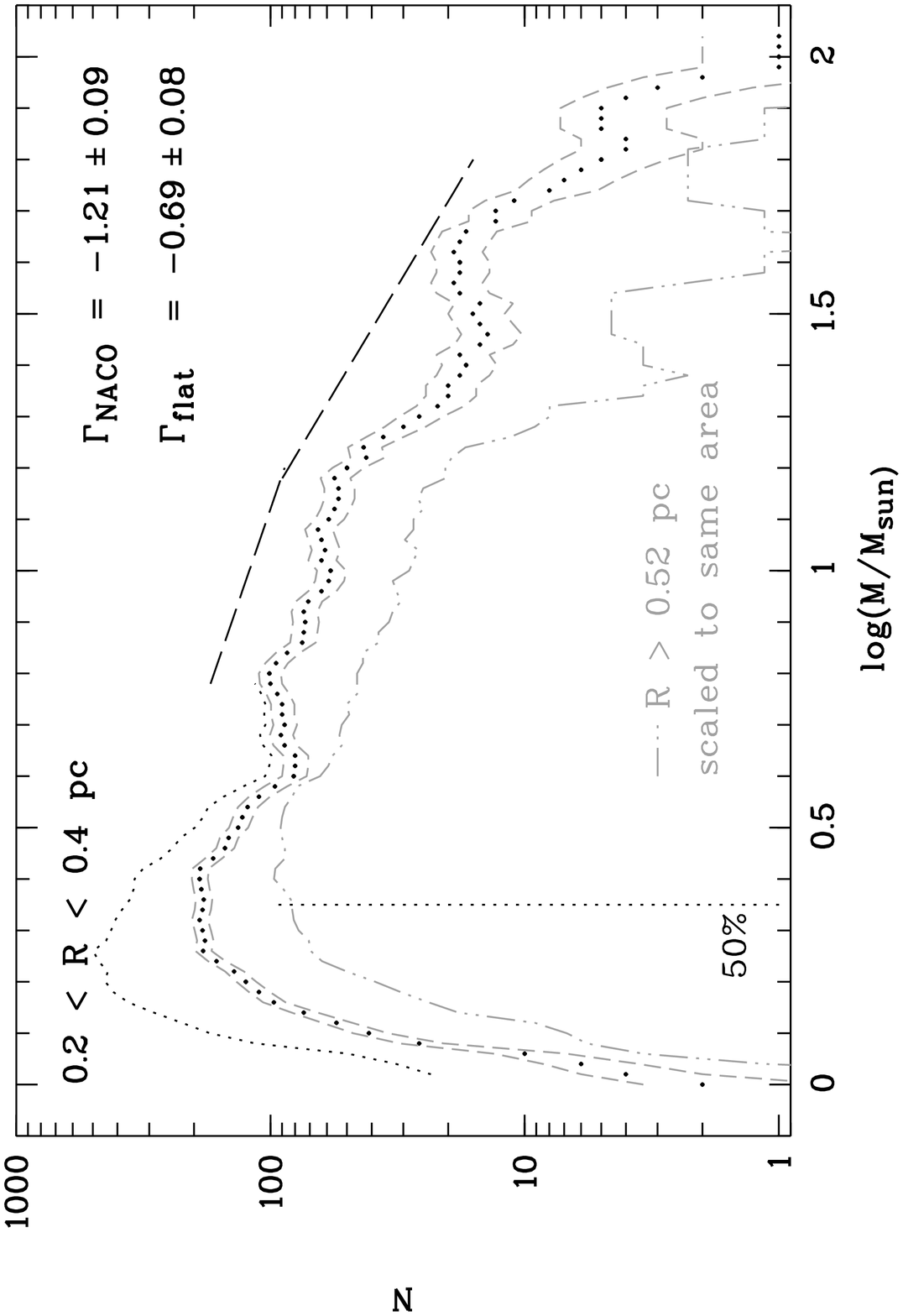}
\caption{\label{nacomfcore}
Mass segregation and turn-over in the Arches cluster core. 
The core MF ($r < 0.2$ pc) (left) is heavily biased to high-mass
stars, with a pronounced TO at $\log m = 0.84$ or $7\,M_\odot$.
The linear, least-squares fit (dashed line) shows that
the MF steepens from $\Gamma \sim -0.26$ ($12 < M < 60\,M_\odot$)
to a Salpeter-like slope of $\Gamma = -1.21$ ($16 < M < 60\,M_\odot$)
at the high-mass end and $\Gamma = -0.69$ ($6 < M < 16\,M_\odot$) for 
intermediate masses in the $2^{nd}$ annulus ($0.2 < r < 0.4$ pc), 
while the TO is still present around $6\,M_\odot$.
The area-scaled MF of the outer NACO field ($r > 0.5$ pc)
indicates that the TO becomes more pronounced when field data are 
available. The similar TO mass in both annuli contradicts 
expectations from dynamical segregation and supports
a low-mass depleted IMF in Arches.}
\end{figure*}

In the case of dynamical segregation, the apparent high-mass TO would 
be a feature of the cluster core, where the massive stars accumulate,
while stars with lower masses are ejected. The median mass is 
expected to decrease with distance from the cluster center.
If the median mass is evidenced in the TO, the TO mass 
should also shift towards lower masses at larger radii.
The core ($r < 0.2$ pc) and $2^{nd}$ annulus ($0.2 < r < 0.4$ pc) MFs
are compared in Fig.~\ref{nacomfcore}.
While the core and $2^{nd}$ annulus MFs significantly vary in shape,
the TO is present at $\log m \sim 0.84$ or $7\,M_\odot$ in the core
and $\log m \sim 0.8$ or $6\,M_\odot$ in the $2^{nd}$ annulus, in both cases above 
the 50\% completeness threshold, i.e.~only marginally higher in the core. 
Although the TO appears much more pronounced in the core, 
the TO mass is {\sl not} shifted significantly to lower masses in 
the $2^{nd}$ annulus, as would be expected from dynamical evolution alone.
The constant TO mass in the inner and outer cluster supports the 
suggestion of a heavily low-mass depleted {\sl IMF} in the Arches cluster.

Above $6\,M_\odot$, both MFs differ significantly.
The core MF features a power law with a flat slope, $\Gamma = -0.26$,
indicating a very efficient production of high-mass stars in the Arches core.
The sharp transition towards a flat MF above $10\,M_\odot$ lets us speculate
that a distinct physical process influences the formation of high-mass
stars, argueing for primordial segregation. 
Such a scenario is consistent with the requirement 
of an exceptionally dense environment to form the most massive stars. 
The MF slope in the $2^{nd}$ annulus steepens to a Salpeter power law with 
$\Gamma = -1.21$ for $M > 16\,M_\odot$, while intermediate mass stars
are still overabundant as indicated by a flattened MF with $\Gamma = -0.7$
for $6 < M < 16\,M_\odot$. Before field contamination starts to dominate 
at $4\,M_\odot$, the MF flattens to $\Gamma \sim 0$ below $6\,M_\odot$.

A KS test comparing the incompleteness-corrected cumulative mass functions (CF)
for $M >  6\,M_\odot$ yields a probability of less than $10^{-4}$ that the core
and ``field'' ($r > 13\arcsec$) CFs are drawn from the same distribution, 
and $0.03$ of the core compared to the $2^{nd}$ annulus. The probability of a 
common origin of the $2^{nd}$ annulus and field CF is as high as 0.26, or 26\%. 
The higher values derived for the $2^{nd}$ annulus indicate that the population 
at intermediate radii has some similarity with both the core and the field, 
but resembles the field much more closely. The KS test results support the 
core population to be distinct from mass distributions at all larger radii,
and in view of the flat MF strengthens the high-mass bias/low-mass depletion 
in the Arches core.

Indirect evidence for a low-mass truncated MF
has been found in young, massive clusters in the center of the starburst 
galaxy M82 (Rieke et al.~1993).
From the mass-to-light ratio, a low-mass cut-off at $2-3\,M_\odot$ is 
deduced assuming a normal high-mass IMF \citep{Smith2001}.
Below $3\,M_\odot$, the low-mass MF in the Arches core 
is remarkably consistent with the area-scaled distribution of ``field'' stars
in the cluster halo ($r > 0.5$ pc), possibly indicating the low-mass
limit in the {\sl IMF} in the GC. However, tidal stripping can as well 
be responsible for the re-distribution of low-mass stars into the cluster
outskirts. Such an effect, if at work in the Arches cluster, could also
be at work in M82.

\section{Summary and Discussion}

We report direct evidence for a low-mass depleted 
MF in a near-by starburst cluster. The present-day MF of the Arches 
cluster 
%derived from NACO high-resolution AO imaging 
displays a TO around $6-7\,M_\odot$.
A strong bias to high- and against low-/intermediate-mass stars 
is observed in the cluster core, which can be caused
by rapid dynamical segregation or the selective formation of 
high-mass stars on the cost of intermediate mass members.
Field stars appear to become dominant around a few solar masses,
indicating that proper field star subtraction might cause a decreasing 
low-mass MF for $M < 4\,M_\odot$. The apparent decrease in the low-mass MF 
can either be caused by tidal stripping of stars below 
$6\,M_\odot$ due to tidal disruption of the cluster in the GC tidal field, 
or a truncated {\sl IMF} in the dense GC environment. 

Star formation theories predict that the characteristic mass depends on the 
initial conditions of the molecular cloud core and its environment. 
For a cloud of temperature $T_{cl}$ with an embedded
core $T_{core}$, the critical mass for cloud collapse is the Bonnor-Ebert mass, %\newline
$M_{BE}\approx 0.35(T_{core}/10{\rm K})^2\cdot (n_{H_2}T_{cl}/10^6 {\rm erg/cm^3})^{-1/2}$ 
\citep{Elmegreen2000}. 
In molecular clouds in the solar neighbourhood, typical values of
$n_{H_2}=10^4\,{\rm cm^{-3}}$, $T_{cl}=100$ K and $T_{core}=10$ K
result in a characteristic mass of $\sim 0.3\,M_\odot$.
Star formation in the GC environment takes place under more extreme
conditions. Increased temperatures and densities, turbulent pressure, and 
the influence of magnetic fields (Mouschovias 1991, 
Shu et al.~2004) may alter the outcome of the star formation process. 
In the Central Molecular Zone the average $T_{core}=70$ K and cloud pressures are higher
due to turbulence (Morris \& Serabyn 1996).
Assuming the thermal pressure is enhanced by a factor of 10
yields a Bonnor-Ebert mass of $\sim 5\,M_\odot$. 
Following Shu et al.~2004, the same characteristic mass arises when taking into 
account a magnetic field with $B_0=100 \mu {\rm G}$ (Morris \& Serabyn 1996). 
Hence the TO in the Arches MF can be understood as a natural result 
of the extreme star-forming environment in the GC.

If a low-mass depleted IMF can be confirmed for the Arches cluster,
this implies that a low-mass truncated and/or flattened IMF could be more adequate
in intense star-forming environments, especially when using stellar
synthesis models to derive the underlying mass distribution and star-formation
history of star clusters and galaxies. While the integrated
luminosity will still be determined by high-mass stars, the underlying population
will contribute significantly less to the build-up of a Galactic field population
of low-mass stars surviving throughout a Hubble time. Metal enrichment will 
appear enhanced when compared to the surviving low-mass population after 
a timescale of a few 100 Myr to a few Gyr, as the fraction of high-mass stars 
expected from a standard IMF will underestimate the true number of massive stars 
that formed and evolved in a galaxy. These implications have to be taken into
account in the derivation of the star-forming histories and evolution of
external galaxies, and especially in the extreme star-forming environments
of distant galaxies in the early universe.

The question whether the Arches {\sl IMF} was indeed truncated
below a few solar masses, whether the GC environment caused an unusually
high TO mass as compared to the low-mass TO at $0.2-0.5\,M_\odot$
observed in the Trapezium \citep{Muench2002,Hille1997},
or whether the cluster's dynamical evolution in the GC tidal field can 
entirely account for the present-day MF can only be solved by investigating
a large area around the cluster and detailed field observations at comparably 
high spatial resolution as the NACO data presented here. 

\acknowledgements
%{\sl Acknowledgements:}
AS acknowledges DGDF grant support from the ESO,
and wishes to thank MPIA and Hans-Walter Rix for being an unusually supportive
thesis host. We thank our referee, Simon Portegies Zwart, for suggestions 
that helped clarify and enhance the paper.

\clearpage

%% Figure Captions
%
%% Figure 1 
%
%
%% Figure 2
%
%
%% Figure 3
%
%
%% Figure 4
%


\clearpage

\begin{thebibliography}{}
\bibitem[Brandner et al.~2001]{Brandner2001}
Brandner, W., Grebel, E. K., Barb\'a, R. H., et al. 2001, AJ, 122, 858
%\bibitem[Devillard 1997]{Devillard1997}
%Devillard, N. 1997, The Messenger No 87, 19
\bibitem[Elmegreen 2000]{Elmegreen2000}
Elmegreen, B. G. 2000, MNRAS, 311, L5
\bibitem[Figer et al.~1999]{Figer1999}
Figer, D. F., Kim, S. S., Morris, M., et al. 1999, ApJ, 525, 750
\bibitem[Figer \& Kim 2000]{FigerKim}
Figer, D. F., Kim, S. S. 2002, ASP Conference Series, 263, ed. Michael M. Shara (San Francisco), 287
\bibitem[Figer et al.~2002]{Figer2002}
Figer, D. F., Najarro, F., Gilmore, Diane, et al. 2002, ApJ, 581, 258
%\bibitem[Fruchter \& Hook 2002]{Fruchter2002}
%Fruchter, A. S., Hook, R.N. 2002, PASP, 114, 144
\bibitem[Hillenbrand 1997]{Hille1997}
Hillenbrand, L. A. 1997, AJ, 113, 1733
\bibitem[Kim et al.~2000]{Kim2000}
Kim, S. S., Figer, D. F., Lee, H. M., Morris, M. 2000, ApJ, 545, 301
\bibitem[Lejeune \& Schaerer 2001]{Lejeune2001}
Lejeune, T., Schaerer, D. 2001, A\&A, 366, 538
\bibitem[Lenzen et al.~2003]{Lenzen2003}
Lenzen, R., Hartung, M., Brandner, W., et al. 2003, SPIE 4841, 944
\bibitem[Muench et al.~2002]{Muench2002}
Muench, A. A., Lada, E. A., Lada, C. J., Alves, J. 2002, ApJ, 573, 366
\bibitem[Morris \& Serabyn 1996]{Morris1996}
Morris, M., Serabyn, E. 1996, ARAA, 34, 645 
\bibitem[Mouscovias 1991]{Mouscovias1991}
Mouschovias, T. Ch. 1991, in The Physics of Star Formation and Early Stellar Evolution, ed. C.J. Lada \& N.D. Kylafis (Dordrecht: Kluwer), 61
\bibitem[Najarro et al.~2004]{Najarro2004}
Najarro, F., Figer, D. F., Hillier, D. J., Kudritzki, R. P. 2004, ApJ, 611, L105
\bibitem[Palla \& Stahler 1999]{PS1999}
Palla, F., \& Stahler, F. W. 1999, ApJ, 525, 772 
\bibitem[Portegies Zwart et al.~2002]{PZ2002}
Portegies Zwart, S. F., Makino, J., McMillan, S. L. W., Hut, P. 2002, ApJ, 565, 265
\bibitem[Rieke et al.~1993]{Rieke1993}
Rieke, G. H., Loken, K., Rieke, M. J., Tamblyn, P. 1993, ApJ, 412, 99
\bibitem[Rousset et al.~2003]{Rousset2003}
Rousset, G., Lacombe, F., Puget, P., et al. 2003, SPIE 4839, 140
\bibitem[Shu et al.~2004]{Shu2004}
Shu, F. H., Zhi-Yun, L., Allen, A. 2004, ApJ, 601, 930
\bibitem[Smith \& Gallagher 2001]{Smith2001}
Smith, L. J, Gallagher, J. S. III 2001, MNRAS, 326, 1027
%\bibitem[1987]{Stetson1987}
%Stetson, P. B. 1987, PASP, 99, 191
\bibitem[Stolte et al.~2002]{Stolte2002}
Stolte, A., Grebel, E. K., Brandner, W., Figer, D. F. 2002, A\&A, 394, 459
\bibitem[Stolte 2003]{Stolte2003}
Stolte, A. 2003, PhD thesis {\sl ``Mass functions and mass segregation in young starburst clusters''}, Ruperto-Carola University of Heidelberg, Germany
\end{thebibliography}
\end{document}